\documentclass[12pt,oneside]{amsart}

\usepackage{amsfonts}
\usepackage{amssymb}
\usepackage{amsmath}
\usepackage{amsthm}

\usepackage[dvips]{graphics}
\usepackage{subfigure}
\usepackage{epsfig}
\usepackage{graphicx}
\usepackage{psfrag}
\DeclareGraphicsExtensions{.eps, .ps}

\usepackage{a4}
\usepackage{verbatim}
\usepackage{natbib}
\usepackage{color}
\usepackage{url}

\begin{document}

\title[]{A Statistical Mechanics Description of Environmental Variability in Metabolic Networks}

\author{Jonathan J. Crofts$^{1}$}
\address{$^1$School of Science and Technology, Nottingham Trent University, Nottingham, NG11 8NS, UK}
\author{Ernesto Estrada$^{2}$}
\address{$^2$Department of Mathematics and Statistics, Institute of Complex Systems, University of Strathclyde, Glasgow, UK.}

\date{\today}

\keywords{Complexity, Returnability, Centrality, Metabolic networks}

\begin{abstract}
Many of the chemical reactions that take place within a living cell are irreversible. Due to evolutionary pressures, the number of allowable reactions within these systems are highly constrained and thus the resulting metabolic networks display considerable asymmetry. In this paper, we explore possible evolutionary factors pertaining to the reduced symmetry observed in these networks, and demonstrate the important role environmental variability plays in shaping their structural organization. Interpreting the returnability index as an equilibrium constant for a reaction network in equilibrium with a hypothetical reference system, enables us to quantify the extent to which a metabolic network is in disequilibrium. 
Further, by introducing a new directed centrality measure via an extension of the subgraph centrality metric to directed networks, we are able to characterise individual metabolites
by their participation within metabolic pathways. To demonstrate these ideas, we study $116$ metabolic networks of bacteria. In particular, we find that the equilibrium constant for the metabolic networks decreases significantly in-line with variability in bacterial habitats, supporting the view that environmental variability promotes disequilibrium within these biochemical reaction systems.
\end{abstract}

\maketitle

\section{Introduction}
\label{sec:intro}
The set of biochemical metabolic reactions within a living cell form a network of chemical transformations known as a {\it metabolic network} \citep{buchanan2010networks,estrada2011networkbook}. Metabolic networks share many interesting features with other reaction networks. For instance, the products of a metabolic reaction act as reactants for other reactions, thus forming a complex network of metabolic reactions. The {\it metabolites}, i.e. those chemicals involved in metabolic reactions, are small molecules which are imported/exported and/or synthesised/degraded inside the organism. They can play the role of {\it substrates} (reactants) or {\it products}. As with many other chemical reactions, in metabolic ones there exists another important contributor, playing the role of catalyst: the {\it enzymes}. Last but not least are the {\it cofactors}, which are small molecules that bind enzymes and can enhance or decrease their catalytic activities. This combination of substrates, products, enzymes and cofactors make metabolic reactions more complex than usual chemical reactions. Representing such a vast complexity of metabolic transformations is a very challenging problem. Typically, a reductionist approach is taken whereby one represents the network of substrates and products linked by their direct transformations into a condensed metabolic network \citep{lacroix08metabolic_review}. However, such oversimplification can lead to a number of problems when information concerning specific metabolic pathways or the individual role of metabolites is to be inferred \citep{arita2004smallworld}. A global analysis of metabolic networks is less prone to such kinds of errors. For instance, \citet{parter2007environmental} found that certain global characteristics of metabolic networks, e.g. modularity, correlated with the level of variability within an organisms environment for some $117$ bacterial species. That is, they found that the more variable the environment, the more modular the metabolic network. This study, however, failed to account for an extremely important characteristic of metabolic and other reaction networks. That is, under determined physiological conditions some reactions are {\it irreversible}, i.e. they take place in one of only two possible directions. The reversibility of a metabolic reaction is controlled by the thermodynamics, the kinetics and the stoichiometry of the reaction. 

Here we analyze metabolic networks for $116$ of the bacterial species previously studied by \citet{parter2007environmental}. These species live in a broad range of habitats such as oceans, salt lakes, thermal vents, soil and within hosts. Using a classification of these species according to variability in their environments we investigate the extent to which the reversibility of an organisms metabolic network impacts on its adaptability. That is, we consider the directed network of metabolic reactions in which the directionality of a given reaction is indicated by an arrow from substrate to product. Our method consists of formulating a statistical mechanics approach to derive a hypothetical equilibrium constant between the real metabolic network and one in which all reactions are reversible. The equilibrium constant measures how far from a globally reversible state, i.e one in which all reactions are reversible, a metabolic network is. Calculating the free energy of the system with respect to this hypothetical equilibrium provides a measure of the `energetic cost' of the evolutionary process giving rise to the corresponding metabolic network. Using this approach we conclude that bacteria living in more variable environments display on average a higher `global reversibility' (or returnability) in their reactions. In contrast, the metabolic networks of those living in very specialized media display significantly reduced levels of returnability. Maintaining the reversibility of all possible metabolic reactions comes at an extremely high cost. Thus, increased returnability in metabolic networks can be considered as an extreme evolutionary strategy due to pressures imposed by the potential lack of certain nutrients within highly variable environments. On the other hand, when the media is specialised, a constant stream of nutrients is more readily available, allowing the returnability of the metabolic network to be kept at a minimum thus saving considerable energy during the evolutionary process.

\section{Background}
\label{sec:back}
\begin{figure}
\centering
\subfigure[Physical system]{
\includegraphics[scale=.2]{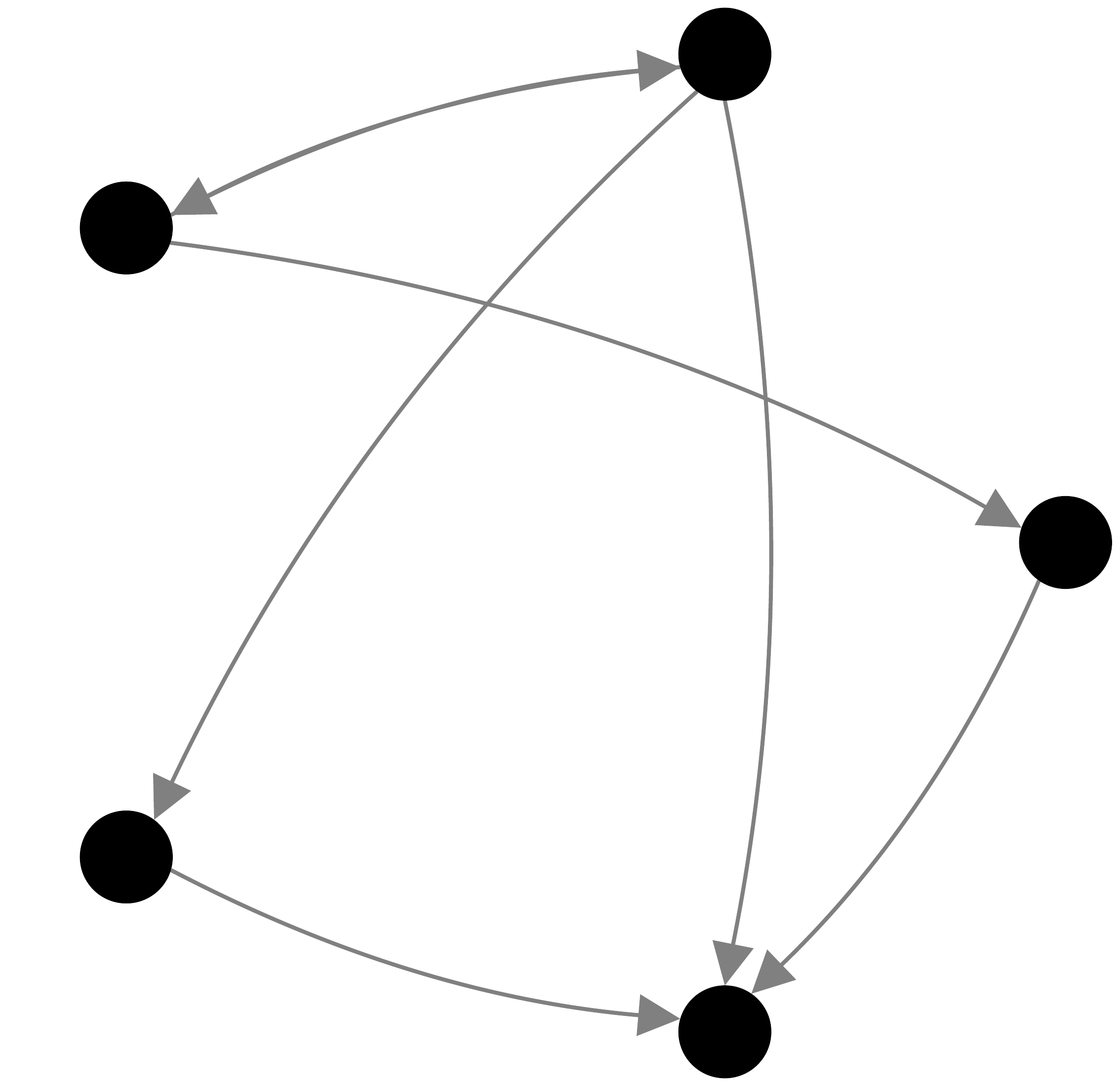}}
\hspace{1cm}
\subfigure[Reference system]{
\includegraphics[scale=.2]{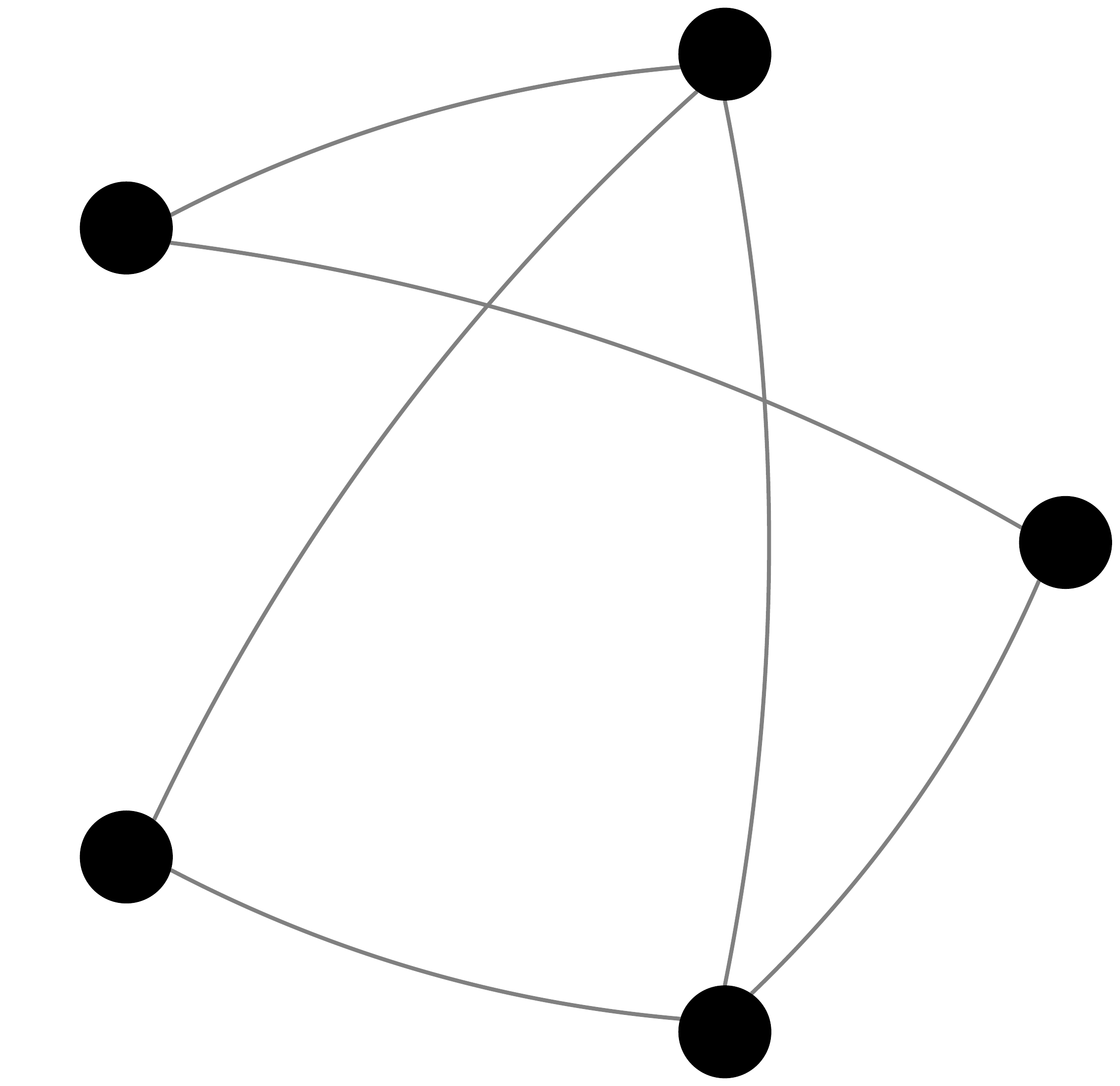}
}
\caption{Reaction network (left) and its reference system (right) consisting of the same set of nodes 
but with every link being undirected.}
\label{fig:1}
\end{figure}

We begin by reformulating the concept of network returnability, as introduced by \citet{estrada2009returnability}, in terms of the equilibrium of a directed reaction graph and its hypothetical reference system. Thus providing a quantitative measure for investigating the extent to which disequilibrium within chemical reaction systems may be considered a product of their natural environments. To start, let us consider that for each reaction network there exists a hypothetical reference system in which all reactions are reversible. In the language of graph theory, the reference system of a set of chemical reactions is nothing other than the underlying, undirected graph for the corresponding reaction graph (see Figure \ref{fig:1}). The reference system considers all reactions to be reversible and thus may be considered as an ideal equilibrium. Note, that a network in which all links are reversible is represented as an undirected graph.

In order to determine how far from this ideal equilibrium a real chemical reaction network lies, we consider the difference in free energy between the physical and reference systems. Let $F_{\mathrm{p}}$ and $F_{\mathrm{r}}$ denote the free energies of the reaction and reference networks respectively. Then
\begin{equation}\label{eqn:partfunc}
\Delta{F} = F_{\mathrm{p}} - F_{\mathrm{r}} = -\beta\ln{\frac{Z_{\mathrm{p}}}{Z_{\mathrm{r}}}},
\end{equation}
where $Z_{\{p,r\}}$ is the partition function of the corresponding (physical or reference) system and $\beta=1/kT$ denotes the inverse temperature\footnote{$k$ is the Boltzmann constant and $T$ the temperature.}. Here temperature may be considered more of a metaphor than a `real' physical parameter, and quantifies the extent to which the network under consideration is subject to external stresses. For instance, it could represent  physiological or extreme environmental conditions under which the metabolic network functions.

Note that the system consisting of only reversible reactions, i.e. the reference system, can be considered as being more stable than the physical system for which only a fraction of reactions are reversible. Now, by considering the  ratio of the two partition functions in (\ref{eqn:partfunc}), we are able to determine the relative returnability \citep{estrada2009returnability} of a reaction network with respect to its hypothetical system
\[
K = \frac{Z(D)}{Z(G)} = e^{-\beta\Delta{F}},
\]
where here, we have used the fact that the chemical reaction system can be represented by the digraph $D = (V, E_D)$ and the corresponding reference system by its underlying graph $G = (V,E_G)$.

Recently, \citet{estrada2007statistical} showed that the partition function of an undirected graph is given by
\begin{equation}\label{eqn:EI}
 Z(A) = \mathrm{trace}\left(e^{\beta{A}}\right),
\end{equation}
where $A$ is the adjacency matrix of the graph. The quantity in (\ref{eqn:EI}) defines the so-called Estrada index of a network \citep{de2007estimating}. Expanding (\ref{eqn:EI}) in a Taylor series gives
\[
 Z(A) = \mathrm{trace}(I) + \beta\mathrm{trace}(A) + \frac{\beta^2}{2!}\mathrm{trace}(A^2) + \cdots + \frac{\beta^k}{k!}\mathrm{trace}(A^k) + \cdots.
\]
Here $I$ denotes the identity matrix. It is well known that the $\mathrm{trace}(A^k)$ counts the number of closed (self-returning) walks of length $k$ within a graph. Thus, $\mathrm{trace}(A^k)>0$ for any network containing at least one cycle of length $k$. The partition function $Z(G)$ can then be considered as a weighted sum over all closed walks (CWs) within the undirected graph -- longer CWs being more heavily penalized by the factor $1/ k!$.

Similarly, one may define the partition function of a directed network as
\begin{align}
 Z(D) &= \mathrm{trace}(e^{\beta D}) = \mathrm{trace}(I) + \beta\mathrm{trace}(D) + \frac{\beta^2}{2!}\mathrm{trace}(D^2) + \cdots\\ \nonumber &+ \frac{\beta^k}{k!}\mathrm{trace}(D^k) + \cdots.
\end{align}
Here, $\mathrm{trace}(D^k)$ counts the number of returnable walks of length $k$ starting and ending at the same node. Again, $\mathrm{trace}(D^k)>0$ if and only if the network has a returnable cycle of length $k$. Now, because $\mathrm{trace}(I)$ is equal to the number of vertices in the graph, and since we are are not interested in the influence network size has on the relative `returnability' of a reaction network, we simply remove this term from the partition functions. Hence, we obtain the following augmented ratio
\[
 K_r(\beta) = \frac{Z(D,\beta)-n}{Z(G,\beta)-n}.
\]
We call the quantity $K_r(\beta)$ the returnability of a directed network. 

The returnability of a directed network is bounded as $0 \leq K_r(\beta)\leq 1$. The lower bound being attained for acyclic digraphs, whilst the upper bound is attained by symmetric digraphs. For the directed cycle on $n$ nodes $C_n$, we have that $K_r\to 0$ as $n\to\infty$ \citep{estrada2009returnability}. Note that for $\beta=0$ the returnability is undefined since $Z(D,\beta)=Z(G,\beta)=0$. This represents the situation in which the network is at infinite temperature with every node in isolation and thus we may set $K_r=0$ in this case. Note, that we define the {\it thermodynamic equilibrium constant} of a network as 
\begin{equation}
pK_r = -\log{K_r}. 
\end{equation}
A large value for this index corresponds to a network that lies far from the ideal equilibrium in which all reactions are reversible, or, in other words, larger values of $pK_r$ are associated  with less returnable networks. 

For completeness, we note that the differences between the returnability metric as defined here and the network reciprocity \citep{garlaschelli2004patterns} are substantial. The latter being the fraction of links in a network that are reciprocal (bidirected). For example, a directed cycle clearly has zero reciprocity yet it is straightforward to realize that the returnability is non-zero in this case. On the other hand, the existence of reciprocal links does not guaranty a high level of returnability within the network. The interested reader is directed to the reference \citep{estrada2009returnability} for further details.

We define the returnability of a given chemical (substrate or product) in a reaction graph analogously to that of the global returnability. That is, given a chemical $i$ we define
\begin{equation}
 K_r(\beta, i) = \frac{(e^{\beta D})_{ii}-1}{(e^{\beta G})_{ii}-1},
 \label{eqn:lret}
\end{equation}
to be the returnability of all chemical reactions in which this chemical takes place. In this way, we may consider Equation (\ref{eqn:lret}) to be a new, directed centrality measure given by the ratio of {\it subgraph centralities} \citep{estrada2005subgraph} for the directed and underlying networks respectively. In the current context, we may interpret this as follows: metabolites that are involved in a larger number of pathways, and particularly shorter, more energy efficient ones, should be considered as being more important to the metabolic process.

\section{Metabolic Networks}
\label{sec:res}

To illustrate our approach, we begin by investigating whether or not a correlation exists between environmental variability and the returnability-equilibrium constant as defined above. To do this, we analyse metabolic networks for some $116$ bacterial species, each of which can be categorised according to their natural environment (see Table \ref{tab:1} for details). The organisms  abide in a variety of habitats, ranging from highly specialised environments, with little, if any, contact with the outside world -- symbiotic bacteria such as {\it buchnera} for example -- to those living in extremely heterogeneous media such as soil, and as such, have evolved under very different selective pressures\footnote{The taxonomy used to rank environmental variability is based upon the NCBI classification for bacterial lifestyle (see \citet{parter2007environmental} and references therein for further details).}. 

\begin{table}[~hbtp]

\begin{center}
\begin{tabular}{|c|ccc|ccc|}
\hline 
Network & \multicolumn{3}{c|}{Nodes} & \multicolumn{3}{c|}{Edges} \\ 
Statistic & \multicolumn{3}{c|}{} & \multicolumn{3}{c|}{} \\ 
& \multicolumn{3}{c|}{\hskip -.15cm min \hskip 0.1cm median \hskip 0.1cm max} & \multicolumn{3}{c|}{\hskip -.15cm min \hskip 0.1cm median \hskip 0.1cm max} \\
\hline 
Obligate (35) & 21 & 89 & 280 & 23 & 102 & 322 \\ 
\hline 
Specialised (5) & 206 & 235 & 299 & 222 & 258 & 336 \\ 
\hline 
Aquatic (4) & 224 & 264 & 386 & 251 & 290 & 440 \\ 
\hline 
Facultative (41) & 40 & 333 & 473 & 42 & 385 & 574 \\ 
\hline 
Multiple (28) & 193 & 454 & 360 & 214 & 431 & 553 \\ 
\hline 
Terrestrial (3) & 282 & 385 & 421 & 329 & 463 & 498 \\ 
\hline 
Total (116) & 21 & 282 & 473 & 23 & 322 & 574 \\ 
\hline 
\end{tabular}\caption{Network statistics for the reaction graphs of the bacterial species studied in this work grouped according to the NCBI classification scheme. According to the NCBI, obligate bacteria have the most constant environment, followed by  specialised and aquatic bacteria, and then facultative,  multiple and terrestrial bacteria in that order (see \citep{parter2007environmental} and references therein for further details).}\label{tab:1}
\end{center}
\end{table}

These metabolic networks are the same as those studied in \citep{parter2007environmental} which were constructed using the KEGG database \citep{kanehisa2000kegg}. Note that, there are a number of ways of representing the set of metabolic reactions of an organism as a simple graph \citep{holme11rsoc}, the most popular of which is a {\it substrate-product graph} whereby each node corresponds to a metabolite, and a connection is made between reacting substances ({substrates) and the products of the reaction. One potential caveat of such an approach, is that it can lead to the erroneous identification of metabolic pathways. Consider, for example, the hypothetical reaction given in Figure \ref{fig:reaction}a. From the corresponding graphical representation (Figure \ref{fig:reaction}b) one may identify a path of length $1$ leading from ATP to Orthophosphate; however, any chemist would immediately object, noting that such a reaction is not possible as ATP must react with H$_2$O in order to produce Orphophosphate. 

\begin{figure}[t]
 \centering{
 \includegraphics[scale=0.75]{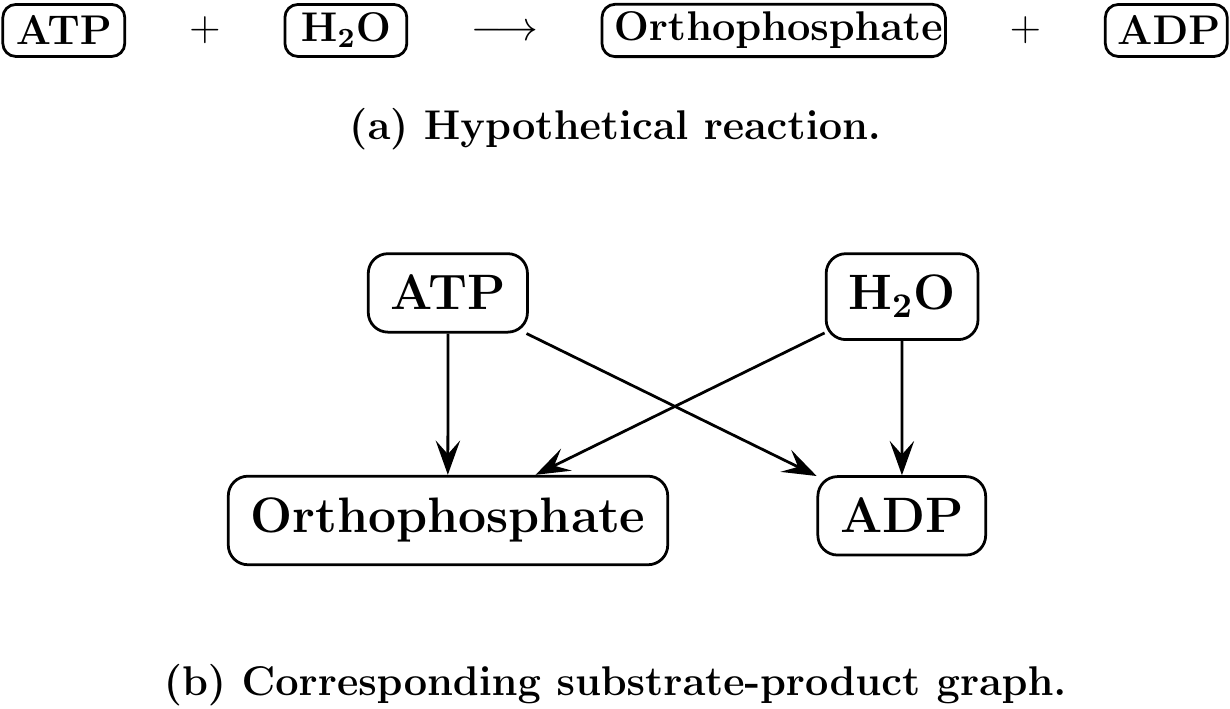}
 }\caption{Illustration of a hypothetical reaction and the corresponding representation as a substrate-product graph.} \label{fig:reaction}
\end{figure}

Much of the criticism of network based studies of metabolism to date are the result of such misrepresentations of the underlying biology \citep{bourguignon2008computational,montanez10}. We note that a correct treatment that avoids such issues is provided via directed hypergraphs \citep{klamt2009hypergraphs,zhou2011hyper}. However, since we are not performing a path analysis and in order to allow for ease of comparison with related studies \citep{parter2007environmental,takemoto2011metabolic,zhou2012community}, we consider the substrate-product representation of metabolism in all of our experiments. Note also, that we remove so-called currency metabolites, e.g., ATP, NADH and H$_2$O, as they tend not to be involved in higher order functions. Additionally, we simplify our analysis by considering only the giant connected component of each reaction graph. 

We calculate the mean thermodynamic index for $116$ of the metabolic networks studied by \citet{parter2007environmental}. In all calculations the inverse temperature was fixed to $\beta=0.25$. Ideally, $\beta$ should be calibrated for each metabolic network according to individual environmental pressures. For example, metabolic networks functioning in extreme conditions should exhibit a value of $\beta\approx 0$ (high temperature) reflecting the increased environmental stresses endured by such systems. However, as the exact conditions for fitting the parameter $\beta$ are unknown, we assume an average value for all environments, selected as that value for which the different bacterial habitats are best discriminated. Figure \ref{fig:enviro_var}(a) shows a plot of the mean thermodynamic index $\langle pK_r\rangle$ versus environmental variability for the different bacterial species. As can be clearly seen, the highest $\langle pK_r\rangle$ value was obtained for those bacteria in the obligate class, then came the specialised and aquatic bacteria, followed by a steady decrease for faculatitive, multiple and terrestrial classes. The group differences observed in Figure \ref{fig:enviro_var}(a) were found to be significant (P-value: $p<10^{-5}$) according to the Kruskal-Wallis (KW) test.

\begin{figure}[t]
 \centering{
 \includegraphics[width=0.9\textwidth]{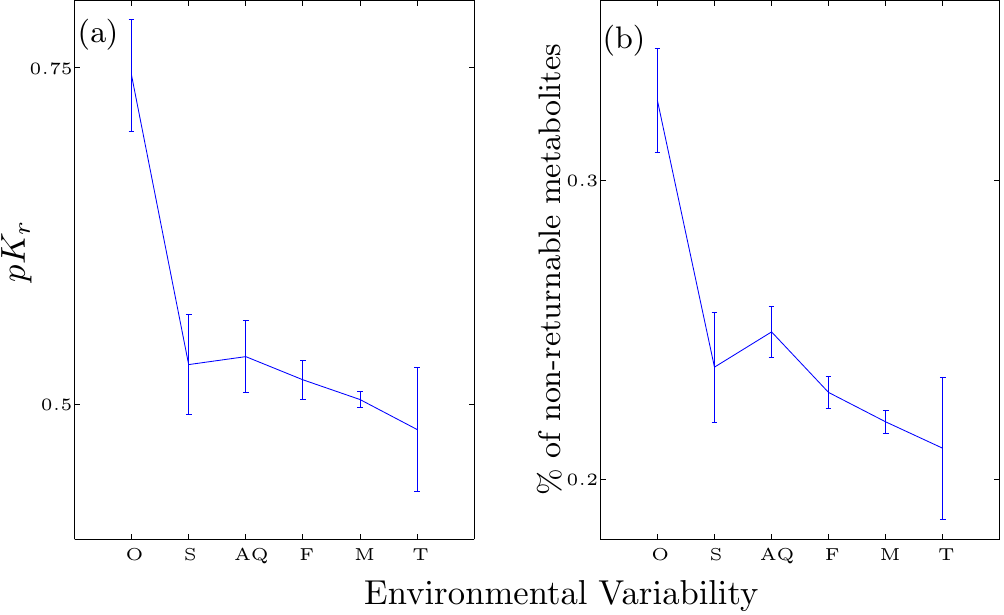}
 }\caption{Relation between environmental variability and (a) the mean thermodynamic equilibrium constant $\langle pK_r(\beta)\rangle$ ($\beta = 0.25$), (b) the mean number of non-returnable metabolites, i.e. those metabolites with zero local returnability $K_r(0.25,i)$. In both cases, the six bacterial habitats under investigation are ordered along the $x$-axis in accordance with the predicted levels of variability provided by \citet{parter2007environmental}: \textbf{O}bligate, \textbf{S}pecialised, \textbf{AQ}uatic, \textbf{F}acultative, \textbf{M}ultiple and \textbf{T}errestrial. Error bars represent standard errors.} \label{fig:enviro_var}
\end{figure}

Additionally, we found that networks from those species residing in more constant environments contained significantly (P-value: $p<10^{-5}$ using the KW test) more non-returnable nodes, i.e. nodes such that $K_r(\beta,i) = 0$. Figure \ref{fig:enviro_var}(b) shows the fraction of non-returnable metabolites versus environmental variability. Note that we can distinguish two subgroups of particular significance within the set of non-returnable metabolites of a reaction graph: (i) the so-called external metabolites, i.e. substrates that are not produced internally by the reaction network; and (ii) those product metabolites that are not fed back into the system. Such metabolites are often referred to respectively as {\it sources} and {\it sinks}. Clearly, we would expect to see a reduction both in the number of external metabolites and the number of products not being consumed via feedback mechanisms as environmental variability increases. 

We remark that this idea relates to previous studies that have found a correlation between genome length and variations in bacterial lifestyle \citep{parter2007environmental,xu2006invited}, in that, constant environments that provide a steady supply of external metabolites lead to redundancy amongst certain genes, which, in turn, leads to genome reduction during the evolutionary process. Further evidence supporting this view was recently provided by \citet{zhou2012community}}, who found that microbes inhabiting varied, heterogeneous environments displayed a larger metabolome than those leading a more specialised lifestyle. For example, \citet{PerezBroca06} have shown that the obligate symbiont {\it Buchnera aphidicola} displays a significantly reduced genome, some $200$ kilobases, when compared against previously sequenced strains. 

\begin{figure}[t]
 \centering{
 \includegraphics[scale = 0.55]{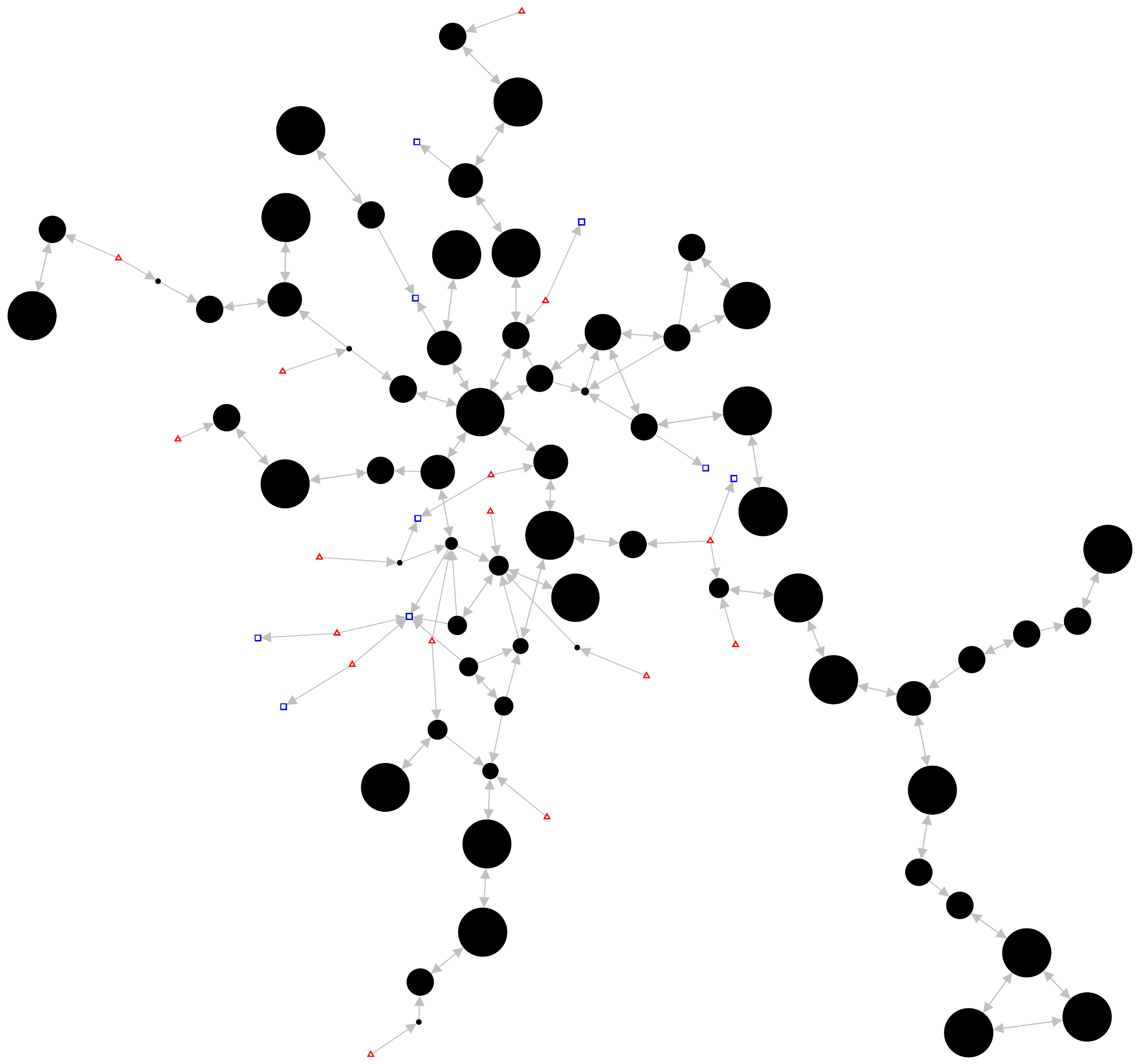}
 }\caption{Visualisation of the metabolic network for the bacterium {\it Buchnera aphidicola}. Nodes are proportional to the local returnability metric given in (\ref{eqn:lret}) with $\beta=0.25$. Source and sink nodes have been highlighted using red triangles and blue squares respectively (colour on-line).} \label{fig:buchnera}
\end{figure}

Figure \ref{fig:buchnera} provides an illustration of the above in the case of {\it Buchnera}. Sink and source nodes are highlighted using red triangles and blue squares respectively (colour on-line); whilst nodes with non-zero local returnability are scaled proportionately. The first interesting observation is the high number of non-returnable nodes, approximately $30\%$ of all metabolites. However, perhaps more interesting is the fact that some $65\%$ of these metabolites are obtained from the environment, via {\it Buchnera-Aphid} symbiosis in this case, thus highlighting the central role played by the host in maintaining functional integrity within these systems.

Note, that in this work we follow the same environmental scheme as the one proposed by \citet{parter2007environmental}. However, when considering our results for specialized and aquatic bacteria we want to make a call for prudence concerning this particular classification scheme. Whilst it is true that some specialised bacteria live in more restricted environments than aquatic ones, Figure \ref{fig:enviro_var} deals with the average thermodynamic equilibrium constant across both specialised and aquatic bacteria. It can clearly be seen that specialized bacteria display a significantly larger variance in their values of $pK_r$ than aquatic ones. The reason for this apparent anomaly can be understood as follows. The specialized bacteria considered in this study are all so-called thermophiles, i.e. {\it A. aeolicus}, {\it C. tepidum}, {\it T. Tengcongensis}, {\it T. Elongatus} and {\it T. Maritima}. However, these bacteria vary significantly with respect to several other environmental factors. For instance, {\it A. aeolicus} requires oxygen to survive whilst {\it C. tepidum} is an anaerobic phototrophic bacterium. Habitats for this last bacterium are quite variable including anoxic and sulfide-rich waters, mud, sediments, microbial mats, and even microbial consortia. The range of temperatures in which these bacteria live also varies significantly, ranging from an optimum growth temperature of $55\,^{\circ}\mathrm{C}$ for {\it T. Elongatus} to water temperature of $80\,^{\circ}\mathrm{C}$ for {\it T. Maritima}. On the other hand, the aquatic bacteria studied here ({\it C. crescentus}, {\it Prochlorococcus}, {\it Synechococcus} and {\it Synechocystis} sp.) live in globally less variable environments. The main difference being that {\it C. crescentus} and {\it Synechocystis} sp. live in fresh water environments while the others live in marine ones. However, many species of {\it Synechococcus} have also been observed in freshwaters. In order to obtain a sense of the level of `specialization' of some aquatic bacteria we mention further the fact that {\it Synechococcus} are also found in oceanic areas which are nutrient depleted, such as the central gyres. Altogether, these facts support our results that indicate (i) a lack of any significant differences between $pK_r$ values for specialised and aquatic groups; and (ii) a significantly increased variability in the values of $pK_r$ for specialised bacteria. 

\begin{figure}[t]
 \centering{\hskip -0.5cm
 \includegraphics[width=0.95\textwidth]{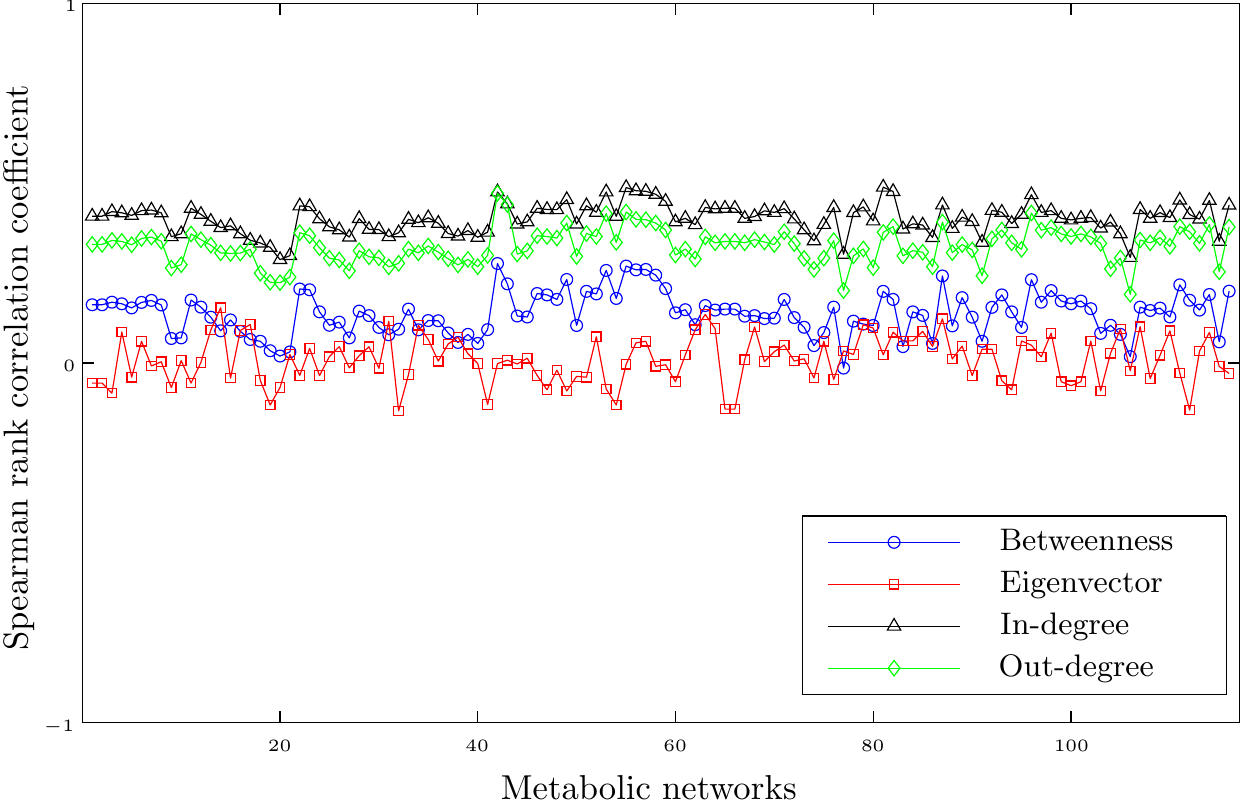}
 }\caption{Spearman rank correlation between the local returnability metric $K_r(\beta,i)$, with $\beta=0.25$, and the centrality measures of betweenness, eigenvector, in-degree and out-degree for $116$ bacterial metabolic networks.} 
\label{fig:spearman}
\end{figure}

Next, we analyse the role of individual metabolites by considering the local returnability metric for the $116$ metabolic networks studied above. The local returnabilty is given by Equation (\ref{eqn:lret}) and represents a new type of directed centrality measure. First, we analyse the extent to which the information contained in this new descriptor is reproduced by other well-known centrality measures. In Figure \ref{fig:spearman} we plot the Spearman rank correlation coefficients between the local returnability and the betweenness, eigenvector, in- and out-degree centralities. In all cases we have considered only directed versions of these network measures. As can be clearly seen, the correlations are low indicating that the information contained in the local returnability metric is not reproduced by any of the other centrality measures. In particular, the eigenvector centrality shows extremely poor correlations, indicating that the local returnability provides a unique spectral measure of centrality.

\begin{figure}[t]
 \centering{\hskip -0.5cm
 \includegraphics[width=0.95\textwidth]{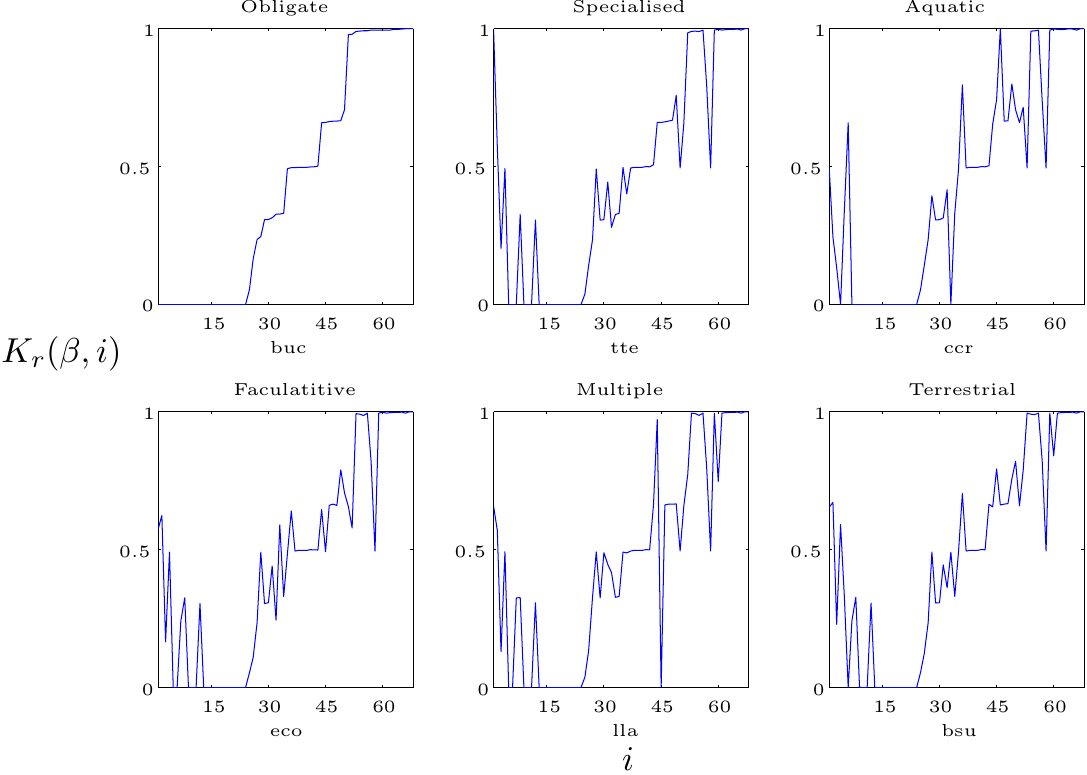}
 }\caption{Local returnability profile for networks from each of the six bacterial lifestyles. Values of $K_r(0.25,i)$ are given for those metabolites that occur in each of the networks ($68$ in total): {\it B. aphidicola} (buc); {\it T. tengcongensis} (tte); {\it C. crescentus} (ccr); {\it E. coli} (eco); {\it L. lactis} (lla) and {\it B. subtilis} (bsu).}  
\label{fig:ret_sig}
\end{figure}

In Figure \ref{fig:ret_sig} we plot the local returnability profiles for illustrative bacterial networks from the six environmental lifestyles. Only metabolites that occur across all six organisms ($68$ in total) are displayed; in all cases, metabolites are ordered   according to the ranking provided by {\it Buchnera}. As can be seen there are a number of universal characteristics shared by these metabolites. Firstly, a small number of metabolites are found to be highly returnable across all six networks, e.g. L-tryptophan, indole and indoleglycerol phosphate. Furthermore, 
a significant overlap, approximately $60\%$, was found between non-returnable metabolites across the six featured networks. In addition to these similarities, a number of substantial differences are readily observed through the plots in Figure \ref{fig:ret_sig}. Perhaps the most striking of which, is the large number of non-returnable metabolites ($24$ of a possible $68$) displayed by the obligate bacterium {\it Buchnera}. Overall, we see that despite their differences, dramatic changes in the profiles of these common metabolites are not present. This can be understood due to the existence of a certain core of common, or universal, metabolites that have evolved in a `standard way' across the different bacteria. Similarity between returnability scores across this common core can be easily calculated using the Pearson correlation coefficient, and we have observed (data not shown) that these correlations are always greater than $0.87$, and in some cases as high as $0.98$, clearly supporting our hypothesis that variability in returnability across these core metabolites is significantly reduced. Thus, the main differences in local and global variability can be attributed to a smaller set of metabolites that are more or less specific to the different kinds of bacteria (according to their environments). Such variations in the local returnability are well described by the values of $K_r(\beta, i)$ which can be further used as an indicator of adaptability and evolution for specific metabolites.

\section{Conclusions}
\label{sec:conc}
In this paper, we considered the difference in free energies, between a
chemical reaction network and its `hypothetical' equilibrium, to be a proxy
for the extent to which a metabolic system is in disequilibrium.  In particular, 
we found that on average `global reversibility' increases significantly
in line with environmental variability, supporting the view that organism
adaptability leads to increased complexities in the resultant metabolic networks. 
In addition, we have proposed a new, directed centrality measure for
characterising nodes (metabolites) in terms of the returnable pathways they
participate.  For the 116 metabolic networks studied here, the new measure
does not show strong correlations with other directed centrality measures,
providing a distinctly different ranking of metabolites.  Moreover, we used
the aforementioned local returnability metric to analyse the role of 
individual metabolites for illustrative networks across the six different lifestyles
studied. We found that common metabolites across these networks were extremely highly correlated, indicating that differences displayed in local and
global returnability may be attributed to a small set of environment specific
metabolites.

\bibliographystyle{agsm}
\bibliography{JTB2013BIB}

\end{document}